\documentclass{aa}           
\def \src {XB\thinspace1916$-$053}
\def \sax {BeppoSAX}
\def\approxgt{\mathrel{\hbox{\rlap{\lower.55ex \hbox {$\sim$}}\kern-.3em \raise.4ex \hbox{$>$}}}}
\input epsf.sty
\begin{document}

   \thesaurus{6(13.25.5;  
               08.09.2;  
               08.14.1;  
               08.02.1;  
               02.01.2)}  
   \title{Progressive covering in dipping and Comptonization in the 
spectrum of XB\thinspace 1916-053 from the BeppoSAX observation}

   \author{M.J. Church\inst{1}, A.N. Parmar\inst{2},
    M. Ba\l uci\'nska-Church\inst{1},
    T. Oosterbroek\inst{2}, D. Dal Fiume\inst{3} and\break M. Orlandini\inst{3}}
   \offprints{M. J. Church}
   \institute{School of Physics and Astronomy, University of Birmingham,
              Birmingham, B15 2TT, UK\\
              email: mjc@star.sr.bham.ac.uk; mbc@star.sr.bham.ac.uk
         \and
             Astrophysics Division, Space Science Department of ESA, ESTEC,
	      Postbus 299, NL-2200 AG Noordwijk, The Netherlands\\
             email: aparmar@astro.estec.esa.nl; toosterb@astro.estec.esa.nl
         \and
             Istituto TESRE, CNR, via Gobetti 101, 40129 Bologna, Italy\\
              email: daniele@tesre.bo.cnr.it; orlandini@tesre.bo.cnr.it}

   \date{Received 27 February 1998; accepted 26 May 1998}
   \authorrunning{M. J. Church et al.}
   \titlerunning{Progressive covering and Comptonization in XB\thinspace 1916-053}

   \maketitle

   \begin{abstract}

We report results of a BeppoSAX observation of the low-mass X-ray
binary (LMXB) dipping source \src. The source joins the small group of LMXB 
detected at energies $\approxgt $100~ keV. The non-dip spectrum is well fitted 
by an absorbed blackbody with a temperature of 1.62$\pm 0.05$~keV
and an absorbed cut-off power law with a photon index of 1.61$\pm 0.01$ and a cut-off
energy of 80$\pm 10$~keV. 
Below 10~keV, where photoelectric absorption is dominant, the dramatic
spectral changes observed during dips can be simply modelled by
progressive covering of the blackbody and cut-off power law components.
The blackbody component is very rapidly absorbed during dips, consistent
with it being point-like, while the cutoff power law is more gradually
absorbed, consistent with it being extended. The most likely locations
for the blackbody component are the surface of the neutron star or the boundary
layer between the neutron star and the accretion disk. The extended 
emission most probably originates in an accretion disk corona. 
Above 10~keV, dipping is detected up to $\sim$40~keV, and there is
some evidence for an energy-independent reduction in
intensity of up to 15\%. This reduction could be caused by electron
scattering or obscuration. In the first case, the change is consistent with an
electron column density of $\sim$$2.9\times 10^{23}$~$\rm {cm^{-2}}$, 
several times smaller than the average hydrogen column 
measured simultaneously.


   \end{abstract}

\section{Introduction}

\src\ is a member of the class of Low Mass X-ray Binaries that exhibit irregular 
reductions, or dips, in X-ray intensity which repeat at the orbital period.
It is generally accepted
that these dips are due to obscuration in the thickened outer regions of a disk where the
accretion flow from the companion star impacts. \src\ has the shortest period
of all dipping sources of 50 min (Walter et al. 1982), and is also notable because of the
difference of $\sim$1\% between the X-ray and optical periods (Grindlay et al. 1988).
The source is a member of the sub-group of dipping sources in which the spectral evolution 
has been fitted by an ``absorbed  plus unabsorbed'' approach, also used for 
XBT\thinspace0748$-$676 and XB\thinspace1254$-$690 (Parmar et al. 1986; Courvoisier et al. 1986). In this
approach, the non-dip spectrum in the band 1--10~keV is fitted by a simple absorbed power-law
or absorbed cut-off power law model. However during dipping intervals, two spectral
components are evident; one absorbed and the other not. It has been customary to fit 
these spectra by dividing the non-dip model into two parts, each having
the power law index of the 
non-dip spectrum, one of which is absorbed, and the other having the non-dip column density. However, 
the normalization of the unabsorbed component decreases markedly (e.g., Smale et al. 1988), and the 
explanation of this is not obvious. It has been suggested that the unabsorbed part of the spectrum 
is due to electron scattering in the absorber. 
 
The dipping LMXB sources do not, in general, show the spectral evolution expected 
for absorption by cold material, ie a hardening of the spectrum as lower energies
are preferentially removed. Some sources show energy independence, eg, X\thinspace1755$-$338  
(White et al. 1984; Church \& Ba\l uci\'nska-Church 1993). 
X\thinspace1624$-$490 shows a strong softening in deep dipping (Church \& 
Ba\l uci\'nska-Church 1995). The absorbed plus unabsorbed sources have strong unabsorbed components. 
This complexity is not predicted by the models where the dips are due to
simple absorption of a point source by neutral material. However it has been
possible to explain the spectral evolution in all of these sources using the model proposed by
Church \& Ba\l uci\'nska-Church (1995), in which the emission consists of point-like blackbody emission 
from the surface of the neutron star, plus an extended Comptonized
component from an accretion disk corona.
In the ASCA observation of \src\, dipping was $\sim$100\%
deep in the 1--10~keV band, showing that all emission components
were removed, and good fits to dip data were obtained
using the two-component model including {\it progressive} covering of the
extended emission region (Church et al. 1997). This
suggests a new explanation for the absorbed and unabsorbed sources in which
a large, dense absorber moves across the source, so that the
covering fraction increases smoothly from zero to unity. The unabsorbed
peak is simply the uncovered part of the extended emission at any stage
of dipping, and with this modelling good fits can be obtained below 10~keV,
without the need for
substantial electron scattering or unusual abundances of the absorbing material. 
Furthermore, calculations of the relative losses 
of X-ray intensity in the absorber due to 
photoelectric absorption and electron scattering showed that, below 10~keV, little scattering is
expected for cosmic abundances (Church et al. 1997).
This is not the case above 10~keV where the Thomson cross
section becomes larger than that of photoelectric absorption.

\src\ has been observed with OSO-8 and {\it Ginga} above 10~keV. From the OSO-8 results
of White \& Swank (1982) it is clear that dipping persisted up to $\sim$20~keV.
In the case of the
{\it Ginga} data (Smale et al. 1992), it is unclear from the published results up to what
energy dipping continued.  \src\ was not detected by OSSE (Barret et al. 1996). 
Until recently, there has been little investigation of LMXB
above 10 keV, although there is evidence from {\it Sigma} results that
the spectra of several low luminosity burst sources extend to high
energies (Barret \& Vedrenne 1994). It is clearly important to investigate
the spectra of LMXB at high energies. In the case of \src\, we show for the first time
that the spectrum extends to above 100 keV, that a Comptonization break
has been seen for the first time in a dipping source, and 
that dipping persists to high energies.

\section{Observations}
\label{sec:observations}

Results from the Low-Energy Concentrator Spectrometer (LECS;
0.1--10~keV; Parmar et al. 1997), Medium-Energy Concentrator
Spectrometer (MECS; 1.3--10~keV; Boella et al. 1997) and the Phoswich
Detection System (PDS; 15--300~keV; Frontera et al. 1997) on-board \sax\
are presented. The MECS consists of three identical grazing incidence
telescopes with imaging gas scintillation proportional counters in
their focal planes. The LECS uses an identical concentrator system as
the MECS, but utilizes an ultra-thin (1.25~$\mu$m) entrance window and
a driftless configuration to extend the low-energy response to
0.1~keV.  The fields of view of the LECS and MECS are circular
with diameters of 37\arcmin and 56\arcmin, respectively. The non-imaging
PDS consists of four independent units arranged in pairs each having a
separate collimator. Each collimator can be alternatively
rocked on- and off-source to monitor the background counting rate.
The hexagonal PDS field of view is 78\arcmin\ full-width at half maximum.

The region of sky containing \src\ was observed by \sax\
between 1997 April 27 21:00 and April 28 19:51~UTC in an
observation lasting 80~ks and spanning 27 orbital cycles.
Good data were selected from intervals when the elevation angle
above the Earth's limb was $>$$4^{\circ}$ and when the instrument
configurations were nominal, using the SAXDAS 1.3.0 data analysis package.
The standard PDS collimator rocking angle of 210\arcmin, and
standard dwell time of 96~s for each on- and off-source position were used.
LECS is operated only during satellite night-time resulting in decreased
exposure. LECS and MECS data were
extracted centered on the position of \src\ using radii of 8\arcmin\ and 
4\arcmin, respectively. Background subtraction in the imaging instruments
was performed using standard files, but is not critical for this source. 
Background subtraction in the PDS was carried out using data from the offset detectors. 
\begin{figure*}[!t]
\epsfxsize=160 mm  
\begin{center}
\leavevmode\epsffile{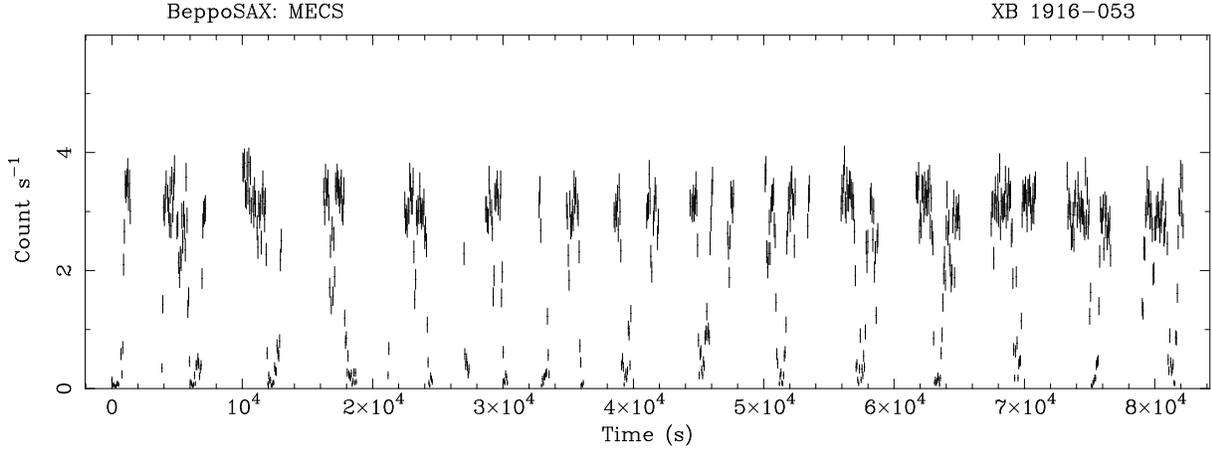}
\end{center}
\caption{Light curve of the complete BeppoSAX observation of \src\
in 64~s bins from merged MECS data in the energy band 1.65--10.0~keV\label{fig1}}
\end{figure*}

\section{Results}
\subsection {Light curves and hardness ratio}

\begin{figure*}[!t]
\epsfxsize=160mm   
\begin{center}
\leavevmode\epsffile{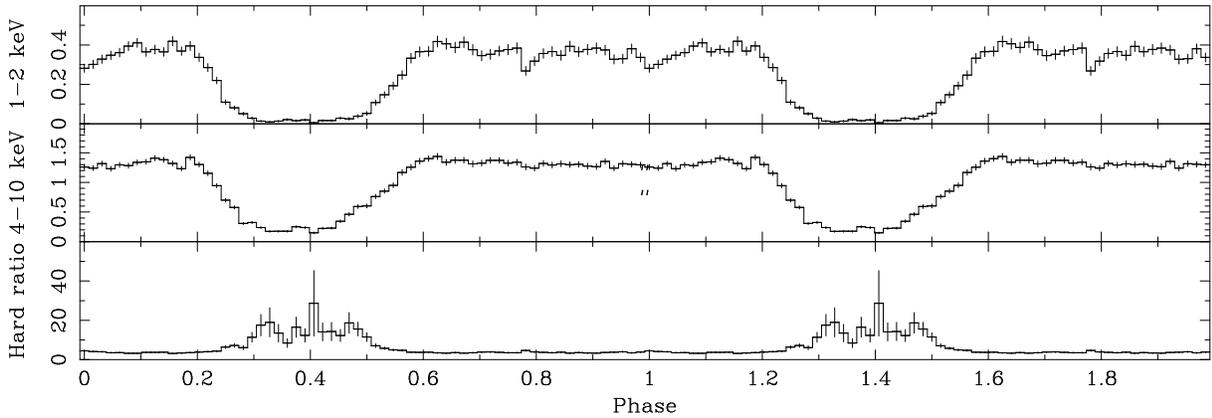}
\end{center}
\caption{Folded MECS light curves: counts~s$^{-1}$ in the bands 1--2~keV and 4--10~keV and
the hardness ratio\label{fig2}}
\end{figure*}

The light curves of all 3 instruments reveal dipping; in the LECS 
and MECS dipping is very strong, however dipping was also 
apparent in the PDS. 
Figure~1 shows the MECS 1.65--10~keV light curve after background subtraction.
Coverage of dipping is very good, with complete
coverage of 11 dips with associated non-dip data. However in the
central part of the observation several dips were only partially observed.
Dipping is very deep, reaching close to 100\% in all 
of the dips observed. This shows that the absorber is both dense and extended 
in relation to all source regions. The reduced exposure of the source in the 
LECS resulted in coverage of non-dip data, dip ingress and deep dipping, but with 
little egress data. 

The best-fit orbital period was obtained using MECS data in which dip
coverage was better.
The data were folded on test periods in the range 2940--3060~s, 
and a best value of 3002$\pm $6~s obtained. The errors are
based on the uncertainty
in the location of the peak of the $\chi^2$ {\it versus} period plot
produced by the period searching program. Because of
the variable nature of the dipping in shape, depth and width, it is
possible that the errors may be underestimated.
The value of 3002$\pm 6$~s is consistent with previous
determinations of 3015$\pm 17$~s ({\it Exosat}, Smale et al. 1989),
3005$\pm 6.6$~s ({\it Ginga}, Smale et al. 1992) and 3005$\pm 20$~s
(ASCA, Church et al. 1997). 

\begin{figure*}[!t]
\epsfxsize=130mm
\begin{center}
\leavevmode\epsffile{fig3.eps}
\end{center}
\caption{The broad-band non-dip spectrum from the LECS, MECS and PDS.
	 The unfolded spectrum (top) shows the individual blackbody, cut-off
	 power law and total spectral components, but in the folded
	 spectrum only the total model is shown\label{fig3}}
\end{figure*}

Figure 2 shows folded background subtracted light-\break curves in the energy ranges
1.0--2.0~keV and 4.0--10.0~keV, together with the hardness ratio
derived from these. 
Dipping is 98\% deep in the band
1.0--2.0~keV, but becoming 100\% at times, and 89\% deep in the 
4.0--10.0~keV energy range. Thus, there is a clear energy dependence of the
dipping, also shown by the width of dipping in the two bands. In the
lower energy range, dipping is very broad and flat-bottomed showing that the
increase in column density is sufficient to reduce the count rate
rapidly to close to zero, which then persists for an interval of 0.2
in phase. In the higher energy band, the count rate continues to fall to
a minimum value at the center of dipping. In both bands, the width
of dipping is 0.39 in phase, equivalent to the absorber subtending
a large angle of 140$^{\circ}$ at the neutron star. Interdipping,
at phases between the main dips,
can be seen in the lower band, and in the raw light curve can be as
deep as 60\%; however it appears to be variable in depth and in phase
and so is smeared out in the folding. The increase in hardness ratio 
in dip ingress and egress show
that part of the spectrum is being removed at low energies, although the
spectrum is complicated by the presence of an unabsorbed peak as discussed
below. 
 
\subsection {The broad-band spectrum}

The spectrum of non-dip emission was investigated by simultaneously fitting
LECS, MECS and PDS data. LECS data were selected with count rate greater
than 1.2~s$^{-1}$, MECS data were selected between 3.1 and 4.0 counts~s$^{-1}$, and
PDS data were selected using the time filters derived from the MECS
selection. 

LECS and MECS data were rebinned to a minimum of 20 counts per channel
giving 392 and 217 channels, respectively. 
PDS data were rebinned to 15 channels spanning the range 13--200~keV. 
Response functions appropriate to the extraction radii in LECS and MECS, 
and to the rebinning in the PDS were used. Responses from Sept. 1997
were used for the MECS and PDS, and the improved response of Feb. 1998
used for LECS.
Factors were applied in the fitting to allow for known normalization differences
between the instruments, the LECS and PDS requiring values of $\sim$0.95 and
$\sim$0.90 relative to the MECS, respectively. LECS data were used in the band 
0.1--10.0~keV, and MECS data between 1.65--10.0~keV,
where the calibration is well established. The photoelectric
absorption coefficients of Morrison \& McCammon (1983) together
with the solar abundances of Anders \& Grevesse (1989) were used
throughout.

\begin{table*}
\caption[]{(a) Parameters of the broadband non-dip spectrum:
$\rm {N_H}$ is in units of $\rm {10^{22}}$ H atom $\rm {cm^{-2}}$,
$\rm {E_{CO}}$ is the cut-off energy;
90\% confidence uncertainties are given. (b) The quality of the fit with
the blackbody component removed}
\halign{#\hfil&&\quad#\hfil\cr
\noalign{\hrule\medskip}
&&$N_{\rm {H}}$&kT (keV) &$\Gamma $&$\rm {E_{CO}}$ (keV) &$\chi^2/\rm {dof}$ \cr
\noalign{\medskip\hrule\medskip}
& a &$0.32 \pm 0.02$& $1.62 \pm 0.05$  &$ 1.61\pm 0.01$& 80.4$\pm $10.0
&537/520\cr
& b &$0.32 \pm 0.02$& \dots                  &$ 1.61       $& 80.4
&2270/525   \cr
\noalign{\medskip\hrule\medskip}}
\end{table*}

\begin{figure*}[!t]
\epsfxsize=110mm   
\begin{center}
\leavevmode\epsffile{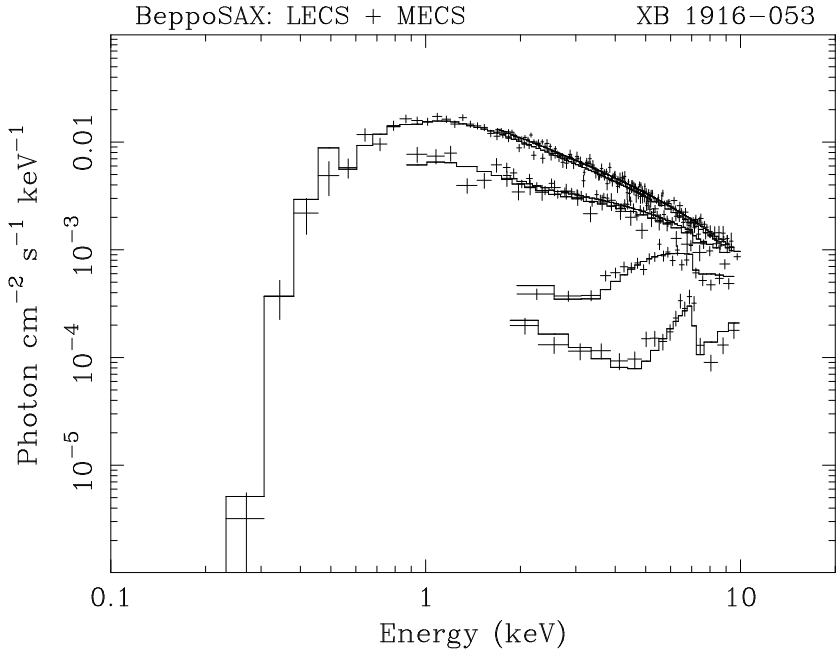}
\end{center}
\caption{Non-dip, intermediate dip and deep dip spectra fitted
by the best-fit models discussed in the text\label{fig4}}
\end{figure*}

Initially, simple models were tested, including an absorbed power law.
There were clear signs of downcurving in the spectrum above
$\sim$35~keV and so channels above this energy were ignored. The fit was
still unacceptable with a $\chi^2$ of 789 for 512 degrees of freedom (dof). 
An absorbed cut-off power law also gave an unacceptable fit, with a lower 
$\chi^2$ of 609 for 522 dof. The spectrum was fitted acceptably below
10~keV, but with large systematic deviations of the model below the data above
30~keV. This was due to the curvature in the spectrum between 1
and 10~keV being modelled by Comptonization down-curving (via a low cut-off
energy $\rm {E_{CO}}$ $\sim$20~keV), clearly incorrect.
A two-component model consisting of a blackbody as well as non-thermal
emission was tried, since this has been previously applied to several dipping sources.
In this previous work, data was not available above 10 keV, and a power
law was used to represent Comptonization at energies $\ll$$\rm {E_{CO}}$, the cut-off
energy. With BeppoSAX, this approximation cannot be used, and
a model of the form: AB*(BB + CPL) was tried, where AB
is an absorption term, and BB and CPL are the blackbody and cut-off power law
components.
This gave a good fit to the non-dip spectrum, with a $\chi^2$ of 537
for 520 dof, and results are shown in Table~1(a) and Fig.~3. 
The blackbody component has an unabsorbed 1--10~keV flux of $\rm
{4.8\times 10^{-11}}$ erg cm$^{-2}$ s$^{-1}$,
19.5\% of the total flux in this band. Compared with the absorbed cut-off
power law model, an F-test indicates that the inclusion of the blackbody is 
significant at $\gg$99.9\% confidence.
A better test of the significance was made by re-fitting the absorbed
cut-off power law, forcing a good fit above 10~keV by fixing $\Gamma $
and $\rm {E_{CO}}$ to the two-component model values. The curvature in the 
spectrum between 1 and 10~keV was badly fitted by the cut-off power law 
without a blackbody, giving a $\chi^2$/dof of 2270/525 as shown in Table 1(b).

We also fitted the LECS and MECS data without
the PDS, to investigate whether using the standard, but relatively
narrow band, stopping at 10 keV affects the spectral fitting results.
Using a cut-off power law model, the value of $\Gamma $ was found
to be $\sim$1.75, with $\rm {kT_{bb}}$ = 1.75~keV and $\chi^2$ =
525 for 505 dof. Fitting a simple power law model, on the basis that
Comptonization down-curving is small $\ll$ the cut-off energy, gives
similar values of $\Gamma $, ie there is little additional
error in using the simple power law. It is thus clear that in a restricted
energy band, there is a tendency to overestimate the power law index,
which is about 10\% in the case of \src\ . The reason
for the effect is probably that there are two sources of curvature in
the spectrum: the blackbody and Comptonization. Without high energy data,
the blackbody contribution may be overestimated, steepening
the measured power law index.

\subsection {Spectral Evolution during dipping}

Dip intervals were selected based on intensity selection of the MECS
data, and time filters corresponding to these selections used to
make LECS spectra. Various intensity bands were tried using up to 7 
contiguous bands, and spectral fitting performed. The final results are 
shown in Fig.~4 with 4 non-contiguous bands only, for clarity. 
A model of the form: AB*BB + AG*PCF*CPL was used, where AG is the non-dip
absorption term and PCF is partial covering,
ie the two-component
model which gave good fits to the non-dip broad-band spectrum, but
allowing progressive covering of the extended emission (see Sect.~1).
It is possible to fit LECS and MECS data simultaneously only for
non-dip and the 1st dip spectrum shown in Fig.~4, since in deeper dipping
insufficient LECS counts are available and only MECS data were fitted.
PDS data would not help clarify the complex changes occurring in the 1--10~keV energy
range. The best-fit parameters of the non-dip spectrum: $\rm
{kT_{bb}}$, $\Gamma $, $\rm {E_{CO}}$ and the normalizations were fixed
in fitting the dip spectra, and good fits were obtained.
Figure 4 shows that the dip spectra clearly display the presence 
of a strong unabsorbed peak which is well-modelled as the uncovered part 
of the extended Comptonized emission at all levels of dipping. 

Difficulty was initially found in fitting the deepest dip spectrum in the
merged MECS data which had been selected in the intensity 
band 0.0--0.45~counts s$^{-1}$.
Poor fits were obtained with values of reduced $\chi ^2$ of $\sim$3. 
When the intensity band was divided into two, strong spectral changes 
were found between these bands, and the lower band
between 0.0--0.3 counts~s$^{-1}$ could be fitted acceptably. 
Figure 4 shows the fit to 4 intensity levels selected in the bands:
0.0--0.3, 0.4--0.8, 1.6--2.2, and 3.1--4.0 counts~s$^{-1}$, and
the corresponding fit parameters are given in Table~2. Figure 5 shows
plots against intensity of the covering fraction, and the blackbody and cut-off
power law column densities for all 7 intensity bands.

\begin{table}
\caption[]{Best fits to the dip spectra shown in Fig. 4:
$\rm {N_H}$ is in units of $\rm {10^{22}}$ H atom $\rm {cm^{-2}}$
for the blackbody and cut-off power-law components; {\it f} is the
covering fraction, and 90\% confidence limits are given}
\halign{#\hfil&&\quad#\hfil\cr
\noalign{\hrule\medskip}
&intensity (count s$^{-1}$) 	&$\rm {N_H^{BB}}$	&$\rm {N_H^{CPL}}$	&f\cr
\noalign{\medskip\hrule\medskip}
&non-dip	& 0.32			&0.32			&0.\cr
&1.6--2.2	&33$\pm 20$	        &7.9$\pm 0.8$		&0.569$\pm 0.020$\cr
&0.4--0.8	&$>$600			&28.4$\pm 1.5$		&0.943$\pm 0.007$ \cr
&0.0--0.3	&$>$600			&105$\pm 5$		&0.977$\pm 0.003$\cr
\noalign{\medskip\hrule\medskip}}
\end{table}

It can be seen that the blackbody column densities increase rapidly, and
have larger values than the Comptonized emission at every level, consistent
with the point source blackbody measuring $\rm {N_H}$ along a localized track, whereas the
$\rm {N_H}$ of the Comptonized emission is an average appropriate
to an extended emission region covered by extended absorber with a
density gradient. Figure 5 shows that the partial covering fraction {\it f}
increases smoothly from zero to close to unity demonstrating that the source
regions are progressively covered by dense absorber, and showing
that the model provides a simple explanation of the complex spectral changes 
observed during dipping.

\begin{figure}[!h]
\epsfxsize=90mm
\begin{center}
\leavevmode\epsffile{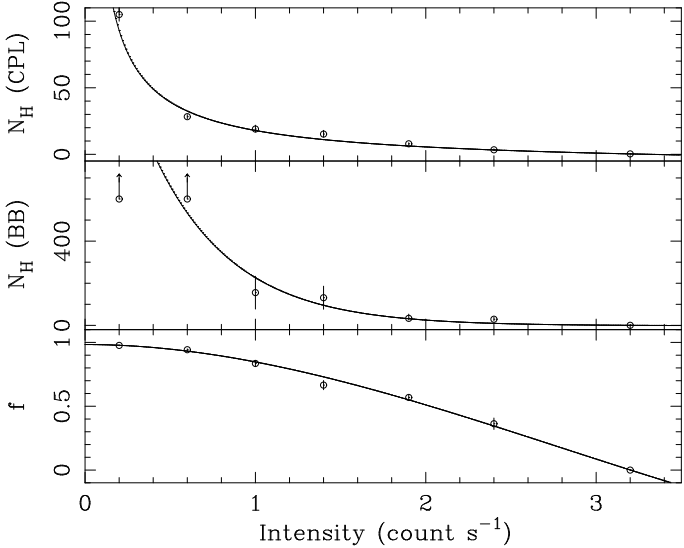}
\end{center}
\caption{The covering fraction {\it f} of the extended non-thermal spectral
component as a function of source intensity, together with the column
densities of the backbody and cut-off power law in units of $\rm
{10^{22}}$ H atom cm$^{-2}$. At the lowest two dip
intensities, the blackbody column densities are lower limits\label{fig5}}
\end{figure}

Next, we compared the non dip and deep dip spectrum using
MECS and PDS data, ie concentrating on the spectrum above 1 keV. 
PDS deep dip data were selected using time filters derived from the MECS, since the
dips are not clearly visible in the PDS.
The main result (Fig.~6) is a systematic reduction in flux in the deep dip PDS
spectrum compared with the non-dip interval. 
The cut-off power law model was fitted to
the non-dip PDS data with the best-fit non-dip parameters. The same model
was applied to the deep dip PDS spectrum, but with the appropriate partial
covering fraction and column density derived from the low-energy fits. 
These parameters do not influence the spectrum above $\sim$20~keV.
Consequently, an 
additional scaling factor was allowed between the deep dip and non-dip
spectra. Without this factor the $\chi^2$ was 27.2 for 13 dof.
The best fit value of the scaling factor was 0.84
giving a $\chi^2$ of 11.3 for 13 dof. Photoelectric absorption will have
effects above 10~keV however, and Fig.~6 does not represent an unambiguous
detection of electron scattering.

\begin{figure}[!t]
\epsfxsize=90mm
\begin{center}
\leavevmode\epsffile{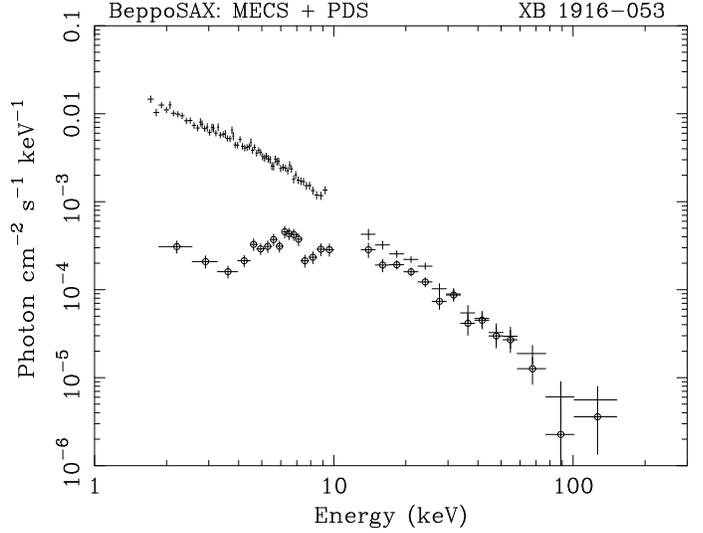}
\end{center}
\caption{Non-dip and deep dip spectra from the MECS1 and PDS instruments
showing dipping persisting to high energies\label{fig6}}
\end{figure}

\section{Discussion}

We have observed the LMXB dipping source \src\ in the very broad
energy range of 0.1--200~keV, using the LECS, MECS and PDS instruments on BeppoSAX.
We find that the spectrum extends to the highest energies of
the PDS, and is a striking example of a LMXB with a strong high energy
component. Moreover, comparison of the non-dip and deep dip spectra reveals 
that dipping is most marked in the 0.1--10~keV range, but also that there are 
significant reductions of flux up to $\sim$40~keV, and some evidence,
although not conclusive, for an energy-independent reduction in PDS flux.
Dipping at high energies must 
be due either to electron scattering or to obscuration of the source
by a very dense absorber. If the dipping is due to
electron scattering, then the reduction factor is $\rm {exp\,-(N_e
\sigma_T)}$ where $\sigma_T $ is the Thomson scattering cross section, 
= $\rm {6.7\times 10^{-25}\; cm^{2}}$. Thus, the average electron
column density $\rm {N_e}$ has the value $\rm {2.6\times 10^{23}\; 
cm^{-2}}$ for a reduction factor of 0.84, 
several times smaller than the largest average value of $\rm
{N_H}$. We can estimate the value of the ionization parameter $\xi$
in the absorbing bulge using the luminosity of the source. During this
observation, the source was relatively faint with 
L$\sim$$\rm {6\times 10^{36}}$ erg $\rm {s^{-1}}$ in the 
energy range 0.5--200~keV for a distance $\sim$9~kpc. 
Using column densities derived at various
levels of dipping, $\xi$ varies from a maximum of
$\sim$50 at the outer edges to $<$1 in the center of the absorber,
implying a low ionization state throughout the absorber. The difference
between $\rm {N_e}$ and $\rm {N_H}$ also indicates that part of the absorber may
be not significantly ionized. 

Next, we find that the non-dip spectrum can be well described 
by the model in which the emission consists of point-source blackbody
emission from the neutron star plus extended Comptonized emission from the
accretion disk corona as proposed by Church \& Ba\l uci\'nska-Church\break (1995). 
This is the first time for this source, and the first time for dipping source in general, 
that the energy band of the instruments includes $\rm {E_{CO}}$.
Values of $\rm {E_{CO}}$ as low as 10~keV have been reported previously,
but must be treated with caution as curvature in the spectrum due to the
blackbody may have been modelled as Comptonization down-curving.
Our observation of down-curving strongly justifies the use of a Comptonizing
term in the two-\break component model; the present work with good fits for progressive covering, and
previous work, show that this is extended emission, and therefore
there is little doubt that it originates in the ADC.
We have found that the cut-off energy is $\sim$80~keV, 
implying an average electron temperature $\rm {kT_e}$ in the Comptonizing
region of $\sim$~30 keV.
From $\Gamma $, we derive a Comptonisation y-parameter of 1.82, and from
the y-parameter $\rm {kT_e}$ relation, an optical depth, $\tau $, of 2.8 and 
hence, an average column density
in the ADC of $\rm {4.2\times 10^{24}}$ H atom $\rm {cm^{-2}}$ are found. Similar
values were obtained by Sunyaev-Titarchuk modelling. 

The spectral fitting results may be used to estimate the sizes of the
emission regions as follows. For the blackbody, the unabsorbed flux in
the band 0.5--200~keV is $\rm {5.5\times 10^{-11}}$ erg cm$^{-2}$
s$^{-1}$, corresponding to a luminosity $\rm {L_{bb}}$ $\sim$ $\rm
{5\times 10^{35}}$ erg s$^{-1}$, and a blackbody radius $\rm {r_{bb}}$ of
0.76~km (defined by $\rm {L_{bb}}$ = $\rm {4\pi\,r^2_{bb}\,\sigma \,T^4}$ where
$\sigma $ is Stefan's constant),
equivalent to 0.5\% of the total surface area of the neutron star. The
diameter, $\rm {d_{ADC}}$, of the extended emission region, the ADC,
can be estimated from the duration $\rm {\Delta t}$ of dip ingress or egress, 
which was found to be $\sim$100~s,
individual values varying between 90 and 150~s. The radius of the accretion
disk will be $\approxgt$ the circularization radius of $\rm {1.9\times
10^{10}}$ cm, and the velocity of the absorbing region 
is given by $\rm {2\,\pi\, r_{disk}/P}$, where $\rm {r_{disk}}$
is the disk radius and P the period, 
and also by $\rm {d_{ADC}/\Delta t}$. From these,
an ADC diameter of $\rm {4.1\times 10^9}$ cm can be derived. Thus the
corona is a region covering the inner 20\% of the accretion disk.

The spectral evolution during
dipping is well described by the two-component model with the
point source blackbody rapidly absorbed, $\rm {N_H}$ increasing to 
$>$$6.0\times
10^{24}$ H atom $\rm {cm^{-2}}$, and the extended Comptonized emission 
covered progressively by the absorber. This method of fitting is able 
to explain dipping in the 1--10~keV energy range as being due primarily to 
photoelectric absorption, without the need to invoke substantial
electron scattering as often proposed in the ``absorbed + unabsorbed'' 
approach. With the present approach, the unabsorbed peak is simply the 
uncovered part of the Comptonized emission as the absorber moves across 
the source.
Above 10~keV the Thomson cross section becomes larger than the
photoelectric absorption cross section, and we may expect some reduction in flux
due to scattering. Our results indicate that an energy-independent decrease
of flux of up to 15\% takes place, but this is not a definite detection of
electron scattering.

\begin{acknowledgements}
The \sax\ satellite is a joint Italian-Dutch programme.
We thank the staff of the \sax\ Science Data Center
for help with these observations. M. J. Church and M. Ba\l uci\'nska-Church
thank the Astrophysics Division of ESTEC for their hospitality during a
recent visit. T. Oosterbroek acknowledges an ESA Fellowship.
\end{acknowledgements}


\begin{thebibliography}{}
\bibitem[1989]{}
Anders E., Grevesse N., 1989, Geochimica et Cosmochimica Acta 53, 197

\bibitem[1993]{}
Barret D., Vedrenne G., 1994, ApJS 92, 505

\bibitem[1996]{}
Barret D., Grindlay J. E., Strickman M. Vedrenne G., 1996, A\&AS 120, 269

\bibitem[1997]{}
Boella G., Chiappetti L., Conti G., et al, 1997, A\&AS 122, 327

\bibitem[1993]{}
Church M.J., Ba\l uci\'nska-Church M., 1993, MNRAS 260, 59

\bibitem[1995]{}
Church M.J., Ba\l uci\'nska-Church M., 1995, A\&A 300, 441

\bibitem[1997]{}
Church M.J., Dotani T., Ba\l uci\'nska-Church M., et al., 1997, ApJ 491, 388

\bibitem[1986]{}
Courvoisier T. J.-L., Parmar A. N., Peacock A., Pakull M., 1986,
ApJ 309, 265

\bibitem[1976]{}
Frontera F., Costa E., Dal Fiume D., et al., 1997, A\&AS 122, 371

\bibitem[1988]{}
Grindlay J.E., Bailyn C.D., Cohn H., et al., 1988, ApJ 334, L25

\bibitem[1976]{}
Morrison D., McCammon D., 1983, ApJ 270, 119

\bibitem[1986]{}
Parmar A.N., White N.E., Giommi P., Gottwald M., 1986,
ApJ 308, 199

\bibitem[1997]{} 
Parmar A.N., Martin D.D.E., Bavdaz M., et al., 1997, A\&AS 122, 309

\bibitem[1988]{}
Smale A.P., Mason K.O., White N.E. Gottwald M., 1988,
MNRAS 232, 647

\bibitem[1989]{}
Smale A.P., Mason K.O., Williams O.R., Watson M.G., 1989, PASJ 41, 607

\bibitem[1992]{}
Smale A.P., Mukai K., Williams O.R., Jones M.H.
Corbet R.H.D., 1992, ApJ 400, 330

\bibitem[1982]{}
Walter F.M., Bowyer S., Mason K.O., et al., 1982, ApJ 253, L67

\bibitem[1982]{}
White N.E., Swank J.H., 1982, ApJ 253, L66

\bibitem[1984]{}
White N.E., Parmar A.N., Sztajno M., et al., 1984, ApJ 284, L9

\end{thebibliography}
\end{document}